\documentclass[12pt,aps]{revtex4}  
  
\usepackage{graphicx}
\usepackage{amsmath} 
\usepackage{textcomp}
\usepackage{mathrsfs}
\usepackage{amsfonts}

\begin{document}
\newcommand{\vett}[1]{\mathbf{#1}}
\newcommand{\uvett}[1]{\hat{\vett{#1}}}
\newcommand{\beq}{\begin{equation}}
\newcommand{\eeq}{\end{equation}}
\newcommand{\barr}{\begin{eqnarray}}
\newcommand{\earr}{\end{eqnarray}}

\title{Cylindrically polarized Bessel-Gauss beams}

\author{Daena Madhi}
\affiliation{$^1$Max Planck Institute for the Science of Light, G$\ddot{u}$nther-Scharowsky-Stra{\ss}e 1/Bau24, 91058 Erlangen, Germany}
\affiliation{$^2$Erlangen Graduate School in Advanced Optical Technologies (SAOT), Paul-Gordan-Stra{\ss}e 6, 91052 Erlangen, Germany}

\author{Marco Ornigotti}
\affiliation{Institute of Applied Physics, Friedrich-Schiller University, Jena, Max-Wien Platz 1, 07743 Jena, Germany}
\email{marco.ornigotti@uni-jena.de}

\author{Andrea Aiello}
\affiliation{$^1$Max Planck Institute for the Science of Light, G$\ddot{u}$nther-Scharowsky-Stra{\ss}e 1/Bau24, 91058 Erlangen, Germany}
\affiliation{$^2$Institute for Optics, Information and Photonics, University of Erlangen-Nuernberg, Staudtstra{\ss}e 7/B2, 91058 Erlangen, Germany}


\begin{abstract}
We present a study of radially and azimuthally polarized Bessel-Gauss beams in both the paraxial and nonparaxial regimes. We discuss the validity of the paraxial approximation and the form of the nonparaxial corrections for Bessel-Gauss beams. We show that, independently from the ratio between the Bessel aperture cone angle $\vartheta_0$ and the Gauss beam divergence $\theta_0$, the nonparaxial corrections are always very small and therefore negligible. Explicit expressions for the nonparaxial vector electric field components are also reported.
\end{abstract}

\pacs{41.20.Jb, 42.25.Ja}
\date{\today}
\maketitle

\section{Introduction}
Cylindrically polarized beams of light, i.e., optical beams whose polarization is non-uniformly distributed across the intensity pattern, have proven to be a very versatile tool because of their peculiar properties such, e.g., the ability of producing a smaller focus \cite{intro2}. Such beams demonstrated to be useful in  various fields of research such as spectroscopy \cite{intro1}, microscopy \cite{intro2.0}, optical tweezing \cite{intro3}, material processing \cite{intro4}, propagation of linear and nonlinear waves in crystals \cite{intro5,intro6,intro7} and quantum information \cite{intro8}. This vast plethora of applications motivated the development of several different experimental techniques to generate such beams \cite{intro9, intro10, intro11, intro12,intro13}. A detailed theoretical analysis of the properties of these beams and their application in the paraxial case can be found in Ref. \cite{Aiello}.

Motivated by these many applications, different groups have then tried in the last years to provide a suitable extension of these beams to the nonparaxial case, by exploring the field of a strongly focused beam \cite{intro14}, using complex dipole sources \cite{intro15}, elegant Laguerre-Gauss beams in the nonparaxial regime \cite{intro16} and vector Bessel beams \cite{intro17}. Recently we also contributed to this subject by proposing a direct and simple generalization of the formalism introduced by Holleczek et al. \cite{Aiello}, based on the use of Bessel beams to generate Hermite-Gaussian-like beams with zero total angular momentum \cite{Ornigotti}. 

Although Bessel beams are exact solutions of the Helmohltz equation, they are  not physical states of the electromagnetic field, as they carry infinite energy \cite{Durnin}. Bessel-Gauss beams, on the other side, are also an exact solutions of the Helmholtz equation, but with a finite energy spectrum \cite{Borghi, April,Greene,Santarsiero}, a feature that makes it possible to realize such beams experimentally \cite{intro18,intro19}. 

It is then the aim of this work to extend the results of Ref. \cite{Ornigotti} to the case of Bessel-Gauss beams by deriving the expressions for the electric field of cylindrically polarized beams of light both in the paraxial and nonparaxial case. Since Bessel-Gauss beams can be nowadays easily generated in an optical laboratory with the help of a suitably programmed spatial light modulator \cite{SLMref, SLMref2}, we believe that the present work could serve as a toolbox to extend the framework of radially and azimuthally polarized states of light to the nonparaxial domain straightforwardly. 

This work is organized as follows: in Sect. 2 we briefly revise the paraxial and nonparaxial form of Bessel-Gauss (BG) beams. These results are then used in Sect. 3 to generate the cylindrically polarized vector fields in the paraxial regime, according to the method presented in Ref. \cite{Ornigotti}. In Sect. 4, we briefly discuss the various regimes of BG beams and how strong is the influence of nonparaxial correction in all these regimes. In Sect. 5, the explicit expression of the vector electric and magnetic fields of cylindrically polarized Bessel-Gauss beams is given. Finally, conclusions are drawn in Sect. 6.

\section{Paraxial and nonparaxial Bessel-Gauss beams}
As it is well known, Bessel beams carry infinite energy, and therefore they do not represent physically meaningful solutions of the Helmholtz equation \cite{Durnin}. This peculiar characteristic is intimately related to the fact that the support of the angular spectrum of such beams is a circle of zero thickness whose radius is given by $K_0=k_0\sin\vartheta_0$ (being $\vartheta_0$ the characteristic cone angle of the Bessel beam) represented by the Dirac delta $\delta(K-K_0)$, a highly singular function. A more realistic description of such beams is represented by Bessel-Gauss beams, that can be thought as the equivalent of Bessel beams where the Dirac-delta circle in Fourier space is replaced by a finite Gaussian distribution. $w_0$ \cite{Borghi}. Another possible interpretation of BG beams is that they are given as a superposition of tilted Gaussian beams with waist $w_0$  whose axes of propagation are uniformly distributed on a surface of a cone of half aperture $\vartheta_0$ \cite{Gori}. In contrast with pure Bessel beams, however, BG beams are not diffractionless anymore, even if they maintain their diffractionless character up to a maximal distance $D= w_{0}/\sin\vartheta_0$ \cite{Borghi}, after which their Gaussian character dominates over the nondiffracting one given by the the Bessel part.  Bessel-Gauss beams are, however, still an \emph{exact} solution of the Helmholtz equation, i.e., 
\begin{equation}\label{eq1}
(\nabla^2+k_0^2)\psi_{\ell}(x,y,z)=0,
\end{equation}
where $k_0=2\pi/\lambda$ is the vacuum wave number. If we write the previous equation in cylindrical coordinates, BG solutions at $z=0$ can be found according to Gori et al. \cite{Gori} to be as follows:
\begin{equation}\label{eq2}
\psi_{\ell}(R,\varphi,0)= J_{\ell}\left(K_{0} R\right)e^{-\frac{{R}^{2}}{w_{0}^{2}}}e^{i\ell\varphi},
\end{equation}
where $K_{0}=k_{0} \sin{\vartheta_{0}}$,  $R=\sqrt{x^{2}+y^{2}}$,  $J_l(x)$ is the Bessel function of the first kind and $(R,\varphi,z)$ are the usual cylindrical coordinates defined with respect to the main axis of propagation $\hat{\textbf{z}}$. The angular spectrum at $z=0$ is then obtained by taking the 2D Fourier transform of Eq. \eqref{eq2}, namely

\begin{eqnarray}\label{eq3}
\tilde{\psi}_{\ell}(K,\phi)&=&\frac{1}{2 \pi} \int d^2R\,\psi_{\ell}(R,\varphi,0) e^{-i\mathbf{K}\cdot\mathbf{R}}\nonumber \\
&=& \frac{w_{0}^{2}}{2  i^{\ell}}  I_{\ell}\left(\frac{w_0^2 K K_{0}}{2}\right) e^{- \frac{w_0^2}{4}\left(K^2+K_0^2\right)} e^{i \ell \phi},
\end{eqnarray}
where $d^2R=dxdy$, $K=\sqrt{k_{x}^{2}+k_{y}^{2}}$,   $ K_{x}=K \cos{\phi}$, $ K_{y}=K \sin{\phi}$, $ \mathbf{R}=x\uvett{x}+y\uvett{y}$ and $I_{\ell}(x)$ is the modified Bessel function of the first kind \cite{nist}. From the previous equation one can easily see that in the limit $w_0\rightarrow\infty$, Eq. \eqref{eq2} gives the traditional Bessel beam, as the Gaussian envelope goes to $1$. Correspondingly,  the angular spectrum defined in Eq. \eqref{eq3} becomes
\begin{equation}\label{eq4}
 \lim_{w_{0} \to \infty} \tilde{\psi}_{\ell}(K, \phi)=\lim_{w_{0} \to \infty}  \left[ w_{0}^{2}\frac{e^{i \ell \phi}}{2 i^{\ell}} I_{\ell}\left(\frac{K K_{0}}{2/w_{0}^{2}}\right) e^{-\frac{K^{2}+K_{0}^2}{2/w_{0}^{2}}}\right]
=\frac{e^{i \ell \phi}}{i^{\ell} K_{0}}\delta(K-K_{0}),
\end{equation}
where in order to calculate the limit we used the following asymptotic expression of the modified Bessel function of the first kind in the vicinity of infinite \cite{nist}:

\begin{equation}
I_{\nu}(z)\simeq\frac{e^z}{\sqrt{2\pi z}}\left(1-\frac{4\nu^2-1^2}{8z}+ \cdots\right),
\end{equation}
Equation \eqref{eq4} is therefore the correct limit that leads to the angular spectrum of a Bessel beams.
To find the expression of the BG beam in the generic plane $z>0$, we now propagate Eq. \eqref{eq3} according to the propagation rule of the angular spectrum \cite{Mandel}, thus obtaining
\begin{eqnarray}\label{eq5}
\psi_{\ell}(R,z)&=&\frac{1}{2 \pi} \int d^2K \,\tilde{\psi}_{\ell}(K,\phi) e^{-i\mathbf{K}\cdot\mathbf{R}} e^{i z \sqrt{k_0^2-K^2}} \nonumber\\
&=&\mathcal{N}\int_0^{\infty}dK\left[ K e^{-\frac{K^2}{4/w_0^2}} I_{\ell} \left( -\frac{w_0^2K K_0}{2}\right) J_{\ell}(K R) e^{i z k_0 \sqrt{1-K^{2}/k_{0}^{2}}}\right],
\end{eqnarray}
where $d^2K=dk_xdk_y$ and $\mathcal{N}=(w_0^2/2)\exp{\left[i \ell \phi-K_0^2/(4/w_0^2)\right]}$. This expression is still exact but cannot be calculated analytically, due to the presence of the square root at the exponent of the last exponential function. However, in the paraxial limit one has that $K/k_0 \ll 1$ and a Taylor expansion of the square root around $K/k_0=0$, i.e., 
\begin{equation}\label{eq6}
 \sqrt{1-K^{2}/k_{0}^{2}}\simeq 1-\frac{1}{2}\left(\frac{K}{k_0}\right)^2+\mathcal{O}\left(\frac{K}{k_0}\right)^4,
\end{equation}
allows us to rewrite the angular spectrum propagator in the approximate form
\beq\label{approx}
e^{ik_0z\sqrt{1-K^2/k_0^2}}\simeq e^{ik_0z}e^{-\frac{izK^2}{2 k_0}},
\eeq
where the quadratic phase factor is the so-called Fresnel propagator and it is responsible for the paraxial propagation \cite{Mandel}. With this in mind, we can now calculate from Eq. \eqref{eq5} the form of the BG beam in the paraxial limit and retrieve the nonparaxial corrections as higher order correction to the paraxial limit. In order to do so, we first need to isolate the Fresnel term from the exact propagator 
\beq
e^{i z k_{0}\sqrt{1-K^{2}/{  k_{0}}}}= e^{i z k_{0}} \exp{\left(-iz \frac{K^2}{2 k_{0}}\right)} \left[\frac{e^{i z k_{0}\sqrt{1-K^{2}/{2 k_{0}}}}}{e^{i z k_{0}}\exp{\left(-iz\frac{K^2}{2 k_{0}}\right)}}\right],
\eeq
and then perform a Taylor expansion of the nonparaxial part of the propagator (the one in square brakets in the previous equation), thus obtaining
\begin{eqnarray}\label{eq8}
 \frac{e^{i z k_{0}\sqrt{1-K^{2}/{ k_{0}}}}}{e^{i z k_{0}}\exp{\left(-iz\frac{K^2}{2 k_{0}}\right)}}&\simeq& 1-\frac{i k_{0}z}{8}\left(\frac{K}{k_{0}}\right)^4-\frac{ik_{0}z}{16}\left(\frac{K}{k_{0}}\right)^6+\cdots.
 \end{eqnarray}
By inserting this result into Eq. \eqref{eq5} we can then write the exact form of the BG beam in a series form as follows:
\begin{eqnarray}\label{eq9}
\psi_{\ell}(R,z)& \simeq&\mathcal{N}\int_0^{\infty}dK\, K e^{-K^2\left(\frac{1}{4/w_{0}^2}+i\frac{z}{2 k_{0}}\right)} I_{\ell} \left( -\frac{w_0^2K K_0}{2/w_{0}^2}\right) J_{\ell}(K R)  \nonumber \\
&\times& \left[1-\frac{ik_{0}z}{8}\left(\frac{K}{k_{0}}\right)^4-\frac{ik_{0}z}{16}\left(\frac{K}{k_{0}}\right)^6...\right] ,\nonumber\\
&=& e^{ik_{0}z}\left[\psi_{\ell}^{(0)}(x,y,z)+\psi_{\ell}^{(1)}(x,y,z)+\psi_{\ell}^{(2)}(x,y,z)+...\right],
\end{eqnarray}
This expression allows us to evaluate all the expansion terms, the lowest one being the paraxial approximation and the higher ones being the nonparaxial corrections.

The paraxial Bessel-Gauss beam is then given by:
\begin{eqnarray}\label{eq10}
\psi_{\ell}^{(0)}(R,\varphi,z)&=&\mathcal{N}\int_0^{\infty}dK\;K e^{-K^2\left(\frac{1}{4/w_{0}^2}+i\frac{z}{2 k_{0}}\right)} I_{\ell} \left( -\frac{w_0^2K K_0}{2/w_{0}^2}\right) J_{\ell}(K R)\nonumber \\
&=&\frac{e^{i \ell \varphi}}{1+i\zeta}\exp{\left[-\frac{1}{1+i\zeta}(\rho^{2}+i\zeta \Theta^{2})\right]}J_{\ell}\left(\frac{2\rho\Theta}{1+i\zeta}\right),
\end{eqnarray}
where $\rho=R/w_{0}$, $\Theta=\sin{\vartheta_{0}/\theta_{0}}$ and $\zeta=z/z_R$, with $w_{0}=\sqrt{2z_R/k_{0}}$ and $\theta_{0}=2/(k_{0} w_{0})$ being the waist and the angular aperture of the beam, respectively. This equation should be compared with Eq. (12) in Ref. \cite{Borghi}: an explicit calculation shows that Eq.(12) in Ref. \cite{Borghi} is incorrect as it does not satisfy the paraxial equation. This is the first result of our paper.
According to Eq. \eqref{eq9}, the first nonparaxial correction can be written in the following simple compact form:
\begin{eqnarray}\label{eq11}
\psi_{\ell}^{(1)}(R,\varphi,z)&=&\frac{-iz\mathcal{N}}{(2k_{0})^3}\int_0^{\infty}dK\;K^5 e^{-K^2\left(\frac{1}{4/w_{0}^2}+i\frac{z}{2 k_{0}}\right)} I_{\ell} \left( -\frac{w_0^2K K_0}{2}\right)J_{\ell}(K R)\nonumber\\
&=& \frac{iz}{2k_{0}} {\partial^2\over\partial z^2} \left[\psi_{\ell}^{(0)}(x,y,z)\right],
\end{eqnarray}
The explicit expression of Eq. \eqref{eq11} evaluated for arbitrary $\ell$ is quite cumbersome and, for the sake of clarity, it will not be reported here. However, in the present work we are interested in the circumstances $\ell=\pm1$ solely and in these cases the formulas are much simpler: 
\begin{eqnarray}\label{eq12}
\psi_{\ell}^{(1)}(R,\varphi,z)\Big|_{\ell=\pm1}&=&\mp\frac{\zeta \theta_{0}^{2}}{(1+i\zeta)^{5}}e^{-\frac{1}{1+i\zeta}(\rho^2+i\zeta \Theta^2)\pm i\varphi}\Big\{\rho \Theta\Big[\frac{3}{2}(1+i\zeta)\nonumber\\
&-&(\rho^2-\Theta^2)\Big] J_{0}\left(\frac{2 \rho \Theta}{1+i\zeta}\right) -\frac{1}{2}\Big[(1+i\zeta)(\rho^2-\Theta^2)\nonumber\\
&-&\frac{1}{2}(\rho^4-6\rho^2\Theta^2+\Theta^4)\Big] J_{1}\left(\frac{2\rho \Theta}{1+i\zeta}\right)\Big\}.
\end{eqnarray}
\section{Cylindrically polarized paraxial Bessel-Gauss beams}\label{section2}
\begin{figure}[!t]
\begin{center}
\includegraphics[width=\textwidth]{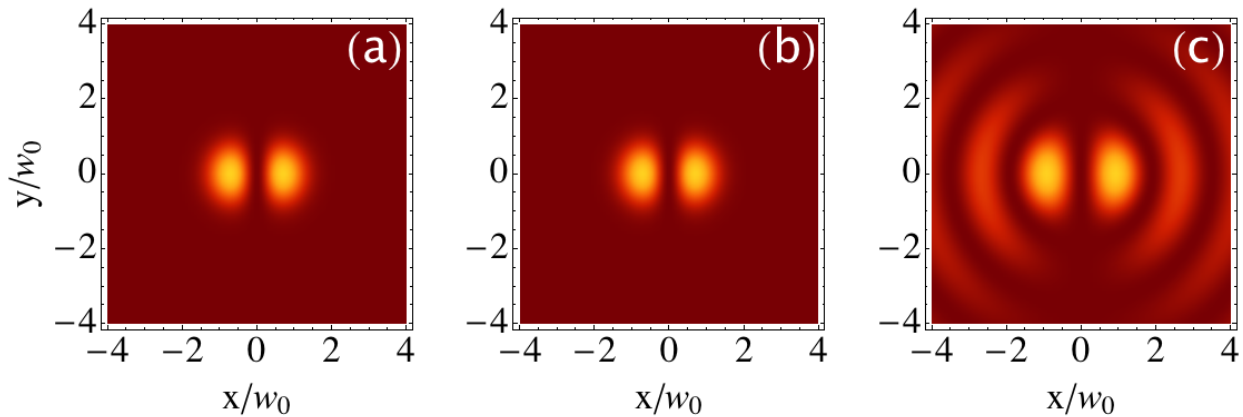}
\caption{(a) Contour plot of the scalar function $\phi_{10}(R,\varphi,0)$ close to the propagation axes. (b)
Contour plot of the Hermite-Gauss beam $HG_{10}(x,y,0)$. (c) Contour plot of the scalar function $\Psi_{10}(R,\varphi,0)$ (as defined in Ref. \cite{Ornigotti}) analogous to $\phi_{10}(R,\varphi,0)$ but defined by using Bessel beams instead of BG beams. A direct comparison between panels (a) and (b) shows that in the
paraxial domain $\phi_{10}$ correctly reproduces the behavior of the Hermite-Gauss beam $HG_{10}$. A comparison between panels (a) and (c) show that the introduction of a Gaussian envelope in the angular spectrum of a BG beam makes the nonparaxial ring to disappear, resulting in a beam carrying finite energy but preserving the same symmetry of $\Psi_{10}$.}
\label{fig1}
\end{center}
\end{figure}

Now that we have correctly calculated the exact form of a paraxial BG beam and its nonparaxial corrections at all orders (each of them can be simply evaluated analytically thanks to the Gaussian form of the integrals), we can now build the Hermite-Gaussian-like BG beams, by combining the paraxial solutions with $\ell=1$ and $\ell=-1$ as follows:
\begin{subequations}\label{eq13}
\begin{eqnarray}
\phi_{10}(R,\varphi,z) &=& \frac{1}{\sqrt{2}}\left[\psi_{1}^{(0)}(R,\varphi,z)+\psi_{-1}^{(0)}(R,\varphi,z)\right]\nonumber\\
&=&\frac{i\sqrt{2}}{1+i\zeta}e^{-\frac{1}{1+i\zeta}(\rho^2+i\zeta\Theta^2)}J_1\left(\frac{2\rho\Theta}{1+i\zeta}\right)\sin\varphi,
\end{eqnarray}
\begin{eqnarray}
\phi_{01}(R,\varphi,z) &=& \frac{-i}{\sqrt{2}}\left[\psi_{1}^{(0)}(R,\varphi,z)-\psi_{-1}^{(0)}(R,\varphi,z)\right]\nonumber\\
&=&-\frac{i\sqrt{2}}{1+i\zeta}e^{-\frac{1}{1+i\zeta}(\rho^2+i\zeta\Theta^2)}J_1\left(\frac{2\rho\Theta}{1+i\zeta}\right)\cos\varphi,
\end{eqnarray}
\end{subequations}
where $\psi_{1}^{(0)}(R,\varphi,z)$ and $ \psi_{-1}^{(0)}(R,\varphi,z)$ are defined by the Eq.\eqref{eq10} for $\ell
=\pm1$ respectively. A sketch of the functon $\phi_{10}(R,\varphi,z)$ in $z=0$ and its comparison with the Hermite-Gaussian beam $HG_{10}(x,y)$ is reported in Fig. \ref{fig1}. As can be noted, the two functions have the same cartesian symmetry. Moreover, Fig. \ref{fig1} also shows that unlike the case of real Bessel beams \cite{Ornigotti} [Fig. \ref{fig1}(c)], BG beams do not present any rings outside the paraxial region. This is a consequence of the fact that their angular spectrum is tailored with a Gaussian function, instead of being a simple Dirac delta function.

\begin{figure}[!t]
\begin{center}
\includegraphics[width=\textwidth]{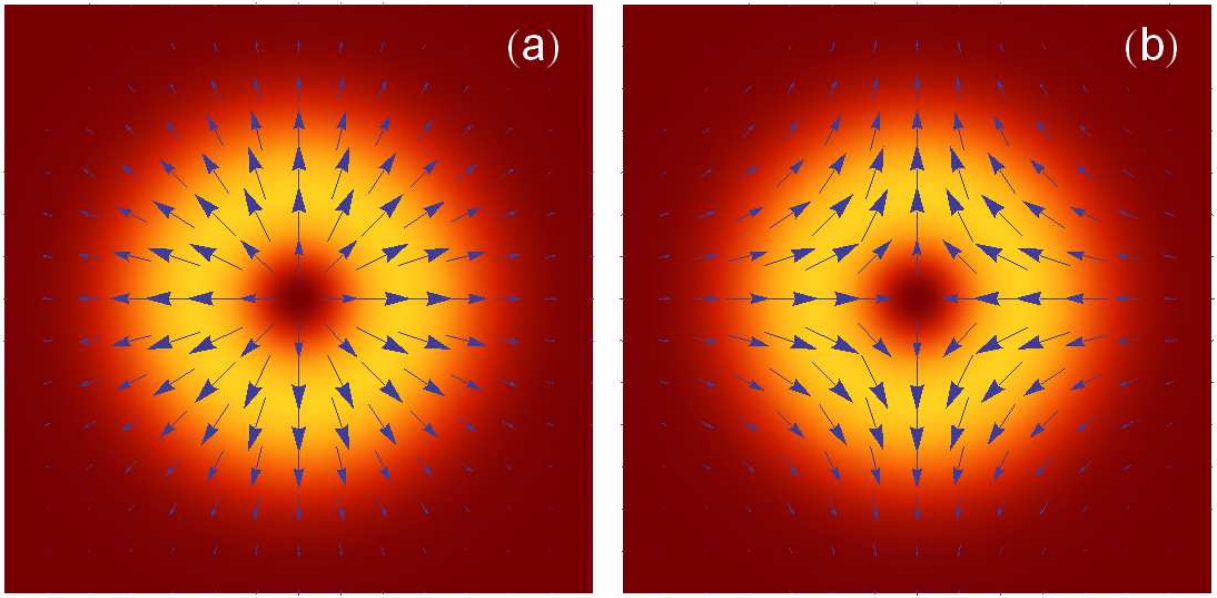}
\caption{Non-uniform polarization patterns of (a) co-rotating radially polarized paraxial mode $\textbf{u}_{R}^+$ and (b) counter-rotating radially polarized paraxial mode $\textbf{u}_{R}^-$, underlayed with the doughnut shaped intensity distribution. The axes of both span the interval [-2,2] in units of beam waist $w_0$.}
\label{fig2}
\end{center}
\end{figure}

\begin{figure}[!t]
\begin{center}
\includegraphics[width=\textwidth]{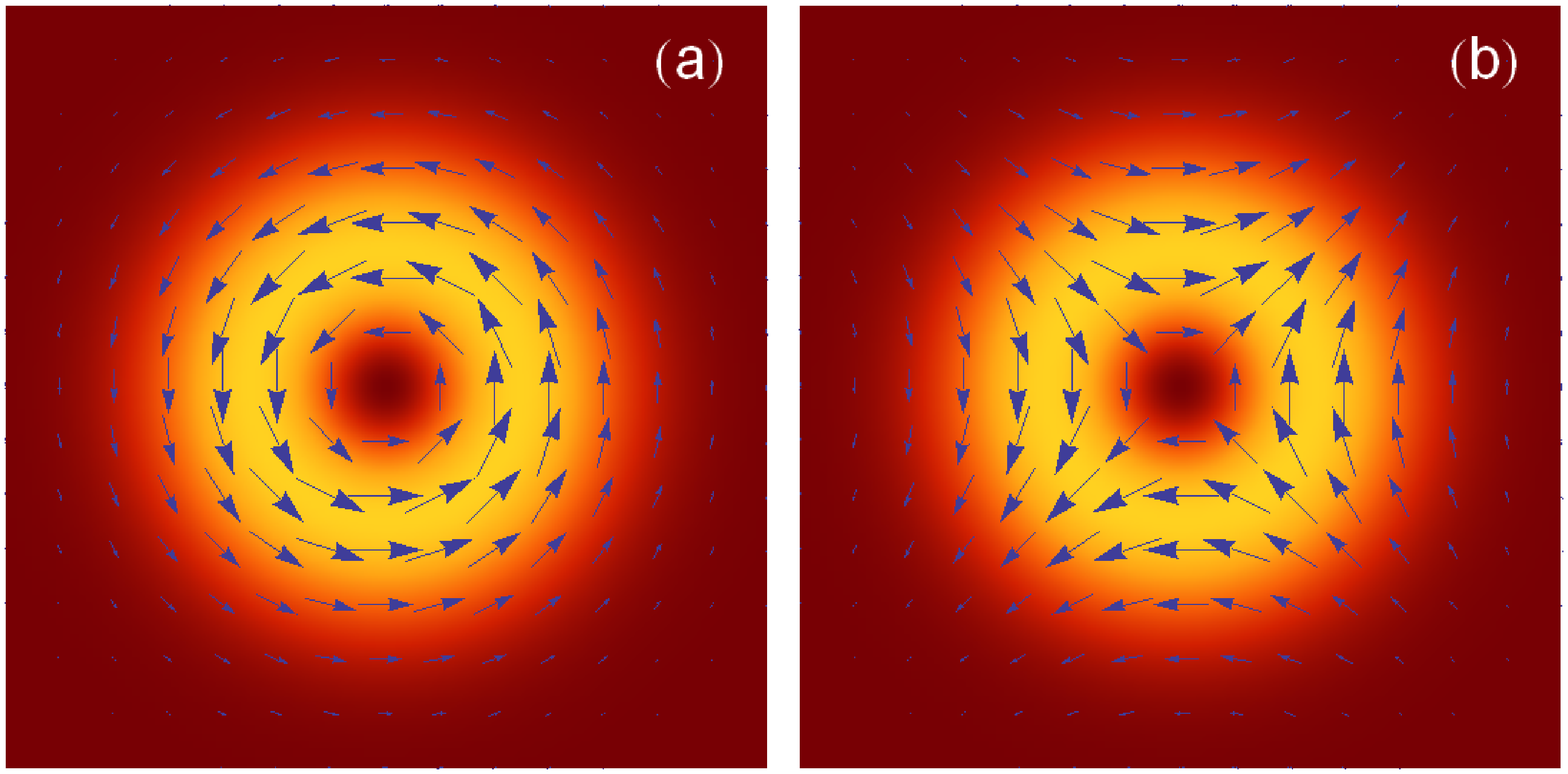}
 \caption{Complex polarization patterns of (a) co-rotating azimuthally polarized paraxial mode $\textbf{u}_{A}^+$ and (b) counter-rotating azimuthally polarized paraxial mode $\textbf{u}_{A}^-$, underlayed with the donut shaped intensity distribution. The axes of both span the interval [-2,2] in units of beam waist $w_0$.}
\label{fig3}
\end{center}
\end{figure}

In analogy with Ref. \cite{Aiello}, we can then build a four dimensional space spanned by the basis formed by the Cartesian product of $\{\phi_{10},\phi_{01}\}$ mode basis defined above and the polarization vectors $\{\hat{\mathbf{x}},\hat{\mathbf{y}}\}$, namely 
\begin{equation}\label{eq14}
\{\phi_{10},\phi_{01}\}\otimes\{\hat{\textbf{x}},\hat{\textbf{y}}\}=\{\phi_{10}\hat{\textbf{x}},\phi_{10}\hat{\textbf{y}},\phi_{01}\hat{\textbf{x}},\phi_{01}\hat{\textbf{y}}\}.
\end{equation}
Radially ($\uvett{u}_R$) and azimuthally ($\uvett{u}_A$) polarized beams can be then easily obtained as linear combinations of these four modes as follows:
\begin{subequations}\label{eq15}
\begin{equation}
\hat{\textbf{u}}_{R}^{\pm} = \frac{1}{\sqrt{2}}(\pm \phi_{10}\hat{\textbf{x}}+ \phi_{01}\hat{\textbf{y}}),
\end{equation}
\begin{equation}
\hat{\textbf{u}}_{A}^{\pm} = \frac{1}{\sqrt{2}}(\mp \phi_{01}\hat{\textbf{x}}+ \phi_{10}\hat{\textbf{y}}),
\end{equation}
\end{subequations}
where the $\pm$ sign refers to co-rotating and counter-rotating modes respectively \cite{Aiello}. The polarization patterns and the intensity profile of these paraxial modes are shown in Fig. \ref{fig2} and \ref{fig3}. 

\section{Nonparaxial corrections}
\begin{figure}[t!]
 \begin{center}
\includegraphics[width=\textwidth]{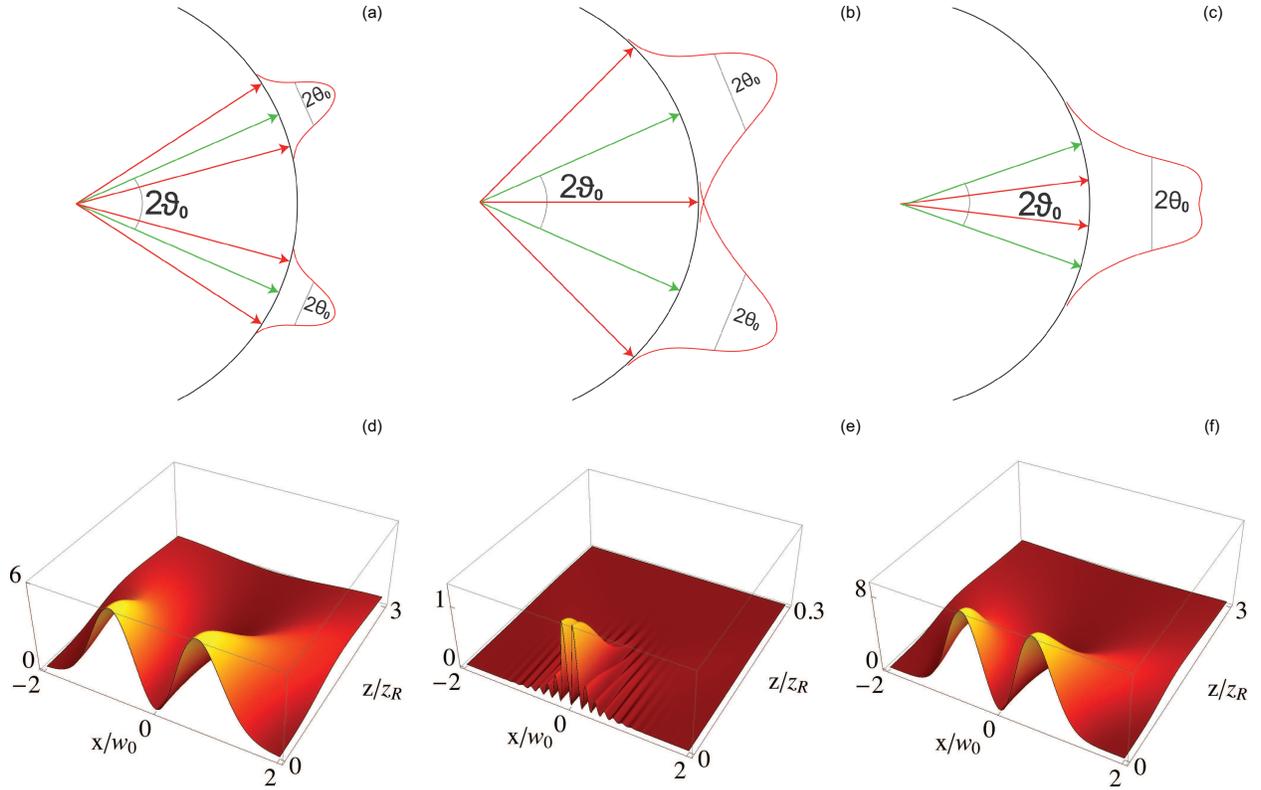}
\caption{Upper row: Schematic representation of the three regimes of BG beams considered in Sect. 4 depending on the ratio between the Bessel cone angle $\vartheta_0$ and the Gaussian beam divergence $\theta_0$. (a)  $\vartheta_0/\theta_0>1$, (b)  $\vartheta_0/\theta_0\simeq 1$ and (c)  $\vartheta_0/\theta_0<1$. Lower row: Three dimensional plot of the $(x,z)$-section intensity of the first nonparaxial correction $\psi_{\ell}^{(1)}(x,0,z)\Big|_{\ell=1}$ as given by Eq. \eqref{eq11} for the three cases (d) $\vartheta_0/\theta_0=10$, (e) $\vartheta_0/\theta_0=1$ and (f) $\vartheta_0/\theta_0=0.1$. In all the figures of the lower panel, the $x$-direction has been normalized to the beam waist $w_0$ while the $z$-direction to the correspondent Rayleigh range $z_R$. The vertical scale, in units of $10^{-6}$, is arbitrary, but the same for all the three plots.}
\label{fig4}
\end{center}
 \end{figure}

A Bessel-Gauss beam is characterized by two competing parameters: the Bessel cone angle $\vartheta_0$ and the width $w_0$ of the Gaussian beam composing the spectrum or, alternatively, its angular spread $\theta_0=2/(k_0w_0)$. Depending on the relative weight of these two parameters, according to Ref. \cite{Gori} we can define three different regimes that are schematically represented in Fig. \ref{fig4}. The first of these regimes corresponds to $\vartheta_0/\theta_0> 1$ [Fig \ref{fig4}(a)]. In this regime, the Gaussian beam components are well separated and the spot size of each single component diffracts during the propagation along $z$. However, up to a distance $D$ defined as the distance from $z=0$ at which a Gaussian beam component has receded from the $z$-axis by a quantity $w_0$ \cite{Gori}, the beam remains diffractionless.

The second regime that we can analyze is given by $\vartheta_0/\theta_0<1$, with $\vartheta_0 \ll 1$. In this case, as it is reported in detail in Ref. \cite{Gori} for the fundamental BG beam, we expect that the central region of the beam (whose radius is approximately $\xi_m/K_0$, being $\xi_m$ the first zero of the function $J_m(\xi)$) closely resembles the central part of a Gaussian beam, as the beam waist $w_0$ of the component gaussian beams is less than the central radius of the BG beam. This correspond to the most paraxial situation. We therefore expect that in this case [Fig. \ref{fig4}(c)] the contribution of the nonparaxial corrections would be negligible. To show this, in Fig. \ref{fig4}(f) we report a section (along the plane $y=0$) of the scalar first order correction $\psi_l^{(1)}(x,z)\Big|_{l=1}$. As can be seen, the intensity of the first nonparaxial order of Eq. \eqref{eq9} is of the order of $10^{-6}$ and it can be therefore neglected.

Although regarding the first case one could intuitively say that the contributions of higher order nonparaxial terms in Eq. \eqref{eq9} are higher than the second one, Fig. \ref{fig4}(d) shows that also in this case the nonparaxial corrections are negligible with respect to the paraxial part of the beam, having an intensity $10^{6}$ times smaller that their paraxial counterpart. 

For the sake of completeness, we present also the intermediate case $\vartheta_0/\theta_0\simeq 1$, where the component Gaussian beams overlap strongly during propagation [Fig. \ref{fig4}(b)]. Also in this case, however, as it appears clear from Fig. \ref{fig4}(e), the effects of the nonparaxial corrections to Eq. \eqref{eq9} are negligible.

In all three regimes,  $\vartheta_0$ and $\theta_0$ are in the paraxial regime. This is why the nonparaxial corrections contribution is negligible. 
\section{Electric and magnetic fields}
The modes obtained from Eqs. \eqref{eq15} and depicted in Figs. \ref{fig2} and \ref{fig3} are strictly paraxial. As already explained in Ref. \cite{Ornigotti}, however,  since 
$\uvett{u}_{R,A}$ are paraxial modes, they are not exact solutions of the Helmholtz equation \eqref{eq1}. In order to fix this problem, in principle, all the nonparaxial corrections to Eq. \eqref{eq11} must be take into account. Once the nonparaxial modes $\vett{U}_{R,A}^{\pm}$ have been calculated by substituting Eq. \eqref{eq9} into the definition  of the Hermite-Gauss-like beams given by Eq. \eqref{eq13} instead of $\psi_{\pm 1}^{(0)}$, they can be used as Hertz vectors to determine the correct form of the electric and magnetic fields, according to the following equations \cite{Ornigotti}:
\begin{eqnarray}\label{eq16}
{\textbf{E}}({\textbf{r}},t) &=& \nabla\times\left[\nabla\times\boldsymbol\Pi(\vett{r},t)\right],\nonumber\\
{\textbf{B}}({\textbf{r}},t) &=& \frac{1}{c^{2}}\frac{\partial}{\partial t}\left[\nabla\times\boldsymbol\Pi(\vett{r},t)\right].
\end{eqnarray}
where $\boldsymbol\Pi({\textbf{r}},t)=\hat{\textbf{U}}_{R,A}^{\pm} \exp(-i\omega t)$ depending on which kind of polarization one wants to attribute to the fields. However, as we discussed in the previous section, the nonparaxial corrections are always very small and they can be neglected irrespectively on the relative weight between the two characterizing parameters of a BG beam, namely $\vartheta_0$ and $\theta_0$. It is therefore sufficient to use the paraxial modes $\vett{u}_{R,A}^{\pm}$ given by Eq. \eqref{eq15} as Hertz potentials to generate the nonparaxial electric and magnetic fields. 
Here we report the explicit expression of the components (in normalized cylindrical coordinates $\{\rho,\phi,\zeta\}$) of the electric field  for all the four cylindrically polarized modes, as deriving from Eq. \eqref{eq16}, whose explicit expression reads as follows:
\begin{subequations}
\begin{align}
E_{R+}^{\rho}(\vett{r},t)&=\frac{i}{(\zeta -i)^5} \Big\{i \left(-2 \rho ^2 \left(i \zeta +3 \Theta ^2+1\right)+\Theta ^2 \left(2 i \zeta +\Theta ^2+2\right)+\rho ^4\right) I_1\left(\frac{2 \Theta  \rho }{-i+\zeta }\right)\nonumber\\
&-2 \Theta  \rho  \left(3 i \zeta +2 \Theta ^2-2 \rho ^2+3\right) I_0\left(\frac{2 \Theta  \rho }{-i+\zeta }\right)\Big\} e^{\chi(\rho,\Theta,\zeta,t) },\\
E_{R+}^{\phi}(\vett{r},t)&=0,\\
E_{R+}^{\zeta}(\vett{r},t)&=\frac{1}{(\zeta -i)^4}\Big\{2 \rho  \left(i \zeta +3 \Theta ^2-\rho ^2+1\right) I_1\left(\frac{2 \Theta  \rho }{-i+\zeta }\right)\nonumber\\
&-2 i \Theta  \left(2 i \zeta +\Theta ^2-3 \rho ^2+2\right) I_0\left(\frac{2 \Theta  \rho }{-i+\zeta }\right)\Big\} e^{\chi(\rho,\Theta,\zeta,t) },
\end{align}
\end{subequations}
for the co-rotating radially polarized electric field,
\begin{subequations}
\begin{align}
E_{R-}^{\rho}(\vett{r},t)&=-\frac{i \cos (2 \phi )}{(\zeta -i)^5 \rho ^2} \Big\{i \Big[2 \rho ^4 \left(-i \zeta -3 \Theta ^2-1\right)+\rho ^2 \left(2 (1+i \zeta ) \Theta ^2-4 i (\zeta -i)^3+\Theta ^4\right)\nonumber\\
&+4 (\zeta -i)^4+\rho ^6\Big] I_1\left(\frac{2 \Theta  \rho }{-i+\zeta }\right)-2 \Theta  \rho  \Big[\rho ^2 \left(3 i \zeta +2 \Theta ^2+3\right)\nonumber\\
&+2 (-1-i \zeta )^3-2 \rho ^4\Big]I_0\left(\frac{2 \Theta  \rho }{-i+\zeta }\right)\Big\} e^{\chi(\rho,\Theta,\zeta,t) }\\
E_{R-}^{\phi}(\vett{r},t)&=-\frac{i \sin (2 \phi )}{(\zeta -i)^5 \rho ^2} \Bigg\{i \Big[2 \rho ^4 \left(-2 \zeta ^2+5 i \zeta +3 \Theta ^2+3\right)+\rho ^2 \Big[2 (-3+\zeta  (2 \zeta -5 i)) \Theta ^2\nonumber\\
&-4 i (\zeta -i)^3-\Theta ^4\Big]+4 (\zeta -i)^4-\rho ^6\Big] I_1\left(\frac{2 \Theta  \rho }{-i+\zeta }\right)\nonumber\\
&+2 \Theta  \rho  \Big[\rho ^2 \left(\zeta  (-4 \zeta +11 i)+2 \Theta ^2+7\right)-2 i (\zeta -i)^3-2 \rho ^4\Big] I_0\left(\frac{2 \Theta  \rho }{-i+\zeta }\right)\Bigg\} e^{\chi(\rho,\Theta,\zeta,t) }\\
E_{R-}^{\zeta}(\vett{r},t)&=\frac{2 \cos (2 \phi )}{(\zeta -i)^4} \Bigg\{\rho  \left(-3 i \zeta -3 \Theta ^2+\rho ^2-3\right) I_1\left(\frac{2 \Theta  \rho }{-i+\zeta }\right)\nonumber\\
&+i \Theta  \left(\Theta ^2-3 \rho ^2\right) I_2\left(\frac{2 \Theta  \rho }{-i+\zeta }\right)\Bigg\} e^{\chi(\rho,\Theta,\zeta,t) }
\end{align}
\end{subequations}
for the counter-rotating radially polarized electric field,
\begin{subequations}
\begin{align}
E_{R-}^{\rho}(\vett{r},t)&=0,\\
E_{R-}^{\phi}(\vett{r},t)&=\frac{i}{(\zeta -i)^5} \Bigg\{i \Bigg[\Theta ^2 \Big[2 \zeta  (-2 \zeta +5 i)-6 \rho ^2+6\Big]\nonumber\\
&+\rho ^2 \left(2 \zeta  (2 \zeta -5 i)+\rho ^2-6\right)+\Theta ^4\Bigg] I_1\left(\frac{2 \Theta  \rho }{-i+\zeta }\right)\nonumber\\
&-2 \Theta  \rho  \Big[\zeta  (-4 \zeta +11 i)+2 \Theta ^2-2 \rho ^2+7\Big] I_0\left(\frac{2 \Theta  \rho }{-i+\zeta }\right)\Bigg\}e^{\chi(\rho,\Theta,\zeta,t) }\\
E_{R-}^{\zeta}(\vett{r},t)&=0,
\end{align}
\end{subequations}
for the co-rotating azimuthally polarized electric field
\begin{subequations}
\begin{align}
E_{A-}^{\rho}(\vett{r},t)&=\frac{\sin (2 \phi )}{(\zeta -i)^5 \rho ^2} \Bigg\{\Bigg[2 \rho ^4 \left(-i \zeta -3 \Theta ^2-1\right)+\rho ^2 \Big[2 (1+i \zeta ) \Theta ^2\nonumber\\
&-4 i (\zeta -i)^3+\Theta ^4\Big]+4 (\zeta -i)^4+\rho ^6\Bigg] I_1\left(\frac{2 \Theta  \rho }{i-\zeta }\right)-2 i \Theta  \rho  \Big[\rho ^2 \left(3 i \zeta +2 \Theta ^2+3\right)\nonumber\\
&+2 (-1-i \zeta )^3-2 \rho ^4\Big] I_0\left(\frac{2 \Theta  \rho }{-i+\zeta }\right)\Bigg\} e^{\chi(\rho,\Theta,\zeta,t) }\\
E_{A-}^{\phi}(\vett{r},t)&=\frac{i \cos (2 \phi )}{(\zeta -i)^5 \rho ^2} \Bigg\{-i \Bigg[2 \rho ^4 \left(-2 \zeta ^2+5 i \zeta +3 \Theta ^2+3\right)\nonumber\\
&+\rho ^2 \Big[2 (-3+\zeta  (2 \zeta -5 i)) \Theta ^2-4 i (\zeta -i)^3-\Theta ^4\Big]\nonumber\\
&+4 (\zeta -i)^4-\rho ^6\Bigg] I_1\left(\frac{2 \Theta  \rho }{-i+\zeta }\right)-2 \Theta  \rho  \Bigg[\rho ^2 \Big[\zeta  (-4 \zeta +11 i)+2 \Theta ^2+7\Big]\nonumber\\
&-2 i (\zeta -i)^3-2 \rho ^4\Bigg] I_0\left(\frac{2 \Theta  \rho }{-i+\zeta }\right)\Bigg\} e^{\chi(\rho,\Theta,\zeta,t) }\\
E_{A-}^{\zeta}(\vett{r},t)&=\frac{2 \sin (2 \phi )}{(\zeta -i)^4} \Bigg\{\rho  \left(3 i \zeta +3 \Theta ^2-\rho ^2+3\right) I_1\left(\frac{2 \Theta  \rho }{-i+\zeta }\right)\nonumber\\
&-i \Theta  \left(\Theta ^2-3 \rho ^2\right) I_2\left(\frac{2 \Theta  \rho }{-i+\zeta }\right)\Bigg\} e^{\chi(\rho,\Theta,\zeta,t) }
\end{align}
\end{subequations}
for the counter-rotating azimuthally polarized electric field. In all these expressions 
\begin{equation}
\chi(\rho,\Theta,\zeta,t)=\frac{-\zeta  \Theta ^2+i \rho ^2}{\zeta -i}-i\omega t,
\end{equation}
 $\vett{r}=\{\rho\uvett{r}+\phi\hat{\boldsymbol{\phi}}+\zeta\uvett{z}\}$ and $I_l(x)$ are the modified Bessel functions of the first kind, which are related with the usual Bessel functions $J_l(x)$ by the relations $I_l(x)=(-i)^lJ_l(ix)$ \cite{nist}. The calculation of the explicit expression of the components of the magnetic field is left to the reader.
\section{Conclusions}
In this work we have theoretically investigated the cylindrically polarized modes associated to Bessel-Gauss beams. We have derived the correct paraxial form of a BG beam in a plane $z\neq 0$ by propagating the angular spectrum and we have expressed the full nonparaxial BG field as a paraxial contribution $\psi_{\ell}^{(0)}(R,\varphi,z)$ plus a series of nonparaxial corrections and we have analyzed their role in three different regimes defined by the ratio $\vartheta_0/\theta_0$. We have shown that independently on the considered regime (corresponding to how much nonparaxial the BG beam is), the nonparaxial corrections are always very small and therefore their contribution can be neglected. 

\section*{Acknowledgements}
DM thanks Prof. Gerd Leuchs for the kind hospitality in his Division at MPL, where this work was made.

\end{document}